# When physics meets biology: a less known Feynman

Marco Di Mauro, Salvatore Esposito, and Adele Naddeo
INFN Sezione di Napoli, Naples - 80126, Italy

We discuss a less known aspect of Feynman's multifaceted scientific work, centered about his interest in molecular biology, which came out around 1959 and lasted for several years. After a quick historical reconstruction about the birth of molecular biology, we focus on Feynman's work on genetics with Robert S. Edgar in the laboratory of Max Delbruck, which was later quoted by Francis Crick and others in relevant papers, as well as in Feynman's lectures given at the Hughes Aircraft Company on biology, organic chemistry and microbiology, whose notes taken by the attendee John Neer are available. An intriguing perspective comes out about one of the most interesting scientists of the XX century.

## 1. INTRODUCTION

Richard P. Feynman has been – no doubt – one of the most intriguing characters of XX century physics (Mehra 1994). As well known to any interested people, this applies not only to his work as a theoretical physicist – ranging from the path integral formulation of quantum mechanics to quantum electrodynamics (granting him the Nobel prize in Physics in 1965), and from helium superfluidity to the parton model in particle physics –, but also to his own life, a number of anecdotes being present in the literature (Mehra 1994; Gleick 1992; Brown and Rigden 1993; Sykes 1994; Gribbin and Gribbin 1997; Leighton 2000; Mlodinov 2003; Feynman 2005; Henderson 2011; Krauss 2001), including his own popular books (Feynman 1985; 1988). If the pictorial representation of Feynman diagrams in quantum field theory is probably his most famous contribution to science (but, certainly, not the only important one), his peculiar life is likely not at all less known to the public due to his involvement in the Manhattan project for the building of the atomic bomb as well as in the panel investigating the Space Shuttle Challenger disaster; physics popularization as well as pedagogical work; political issues and – last but not least – his drum playing and similar extravagant things.

Feynman's genuine interest in the study of Nature often led him to particularly distant areas of research, whose borders were easily crossed by his own curiosity. For example, after the completion of his 1955 work on polaron physics (Feynman 1962), Feynman decided to spend his summer time at Caltech, making excursions into different fields ranging from engineering to biology.

Robert Hellwarth, a research fellow of Feynman at Caltech, moved to Hughes Aircraft Company (1955-1965) and arranged for Feynman to give there lectures for scientists, engineers and technicians on subjects of mutual interest. Feynman continued lecturing regularly at Hughes for many years on a variety of topics, ranging from astrophysics and cosmology to classical and quantum electrodynamics, relativity, scattering theory, as well as mathematical methods in physics and even molecular biology.

Feynman's interest in biology began around 1959, and culminated in the publication of a relevant paper on genetics in 1962 (Edgar *et al.* 1962). His peculiar guiding view was that "there is nothing that living things do that cannot be understood from the point of view that they are made of atoms according to the laws of physics" (Feynman, Leighton and Sands 2005).

The first occasion given to him to reason about such things was probably the talk he delivered at the Annual Meeting of the American Physical Society in December 1969 at Caltech, curiously titled "There's plenty of room at the bottom" (Feynman 1960). Though a popular talk, it is credited as introducing the concept of nanotechnology, since he highlighted the problem of manipulating and controlling things on a small scale. Particularly interesting is Feynman's reasoning about it: "I am inspired by the biological phenomena in which chemical forces are used in a repetitious fashion to produce all kinds of weird effects" (Feynman 1960).

Feynman spent his entire sabbatical year 1959-1960 at Caltech working on biology. With Robert S. Edgar, he worked in the laboratory of Max Delbruck on a project about the characterization of back-mutations, while with Matt Meselson he worked on ribosomes. Given the relevant results he obtained, Feynman was invited to give a seminar on his work at Harvard, where he met James Watson, Francis Crick and others. Interesting enough, a key paper by Crick *et al.* (1961) quoted Feynman's work with Edgar, which was then published in 1962 (Edgar *et al.* 1962).

In the present paper, we dwell just on Feynman's incursions in the field of biology, by focusing on his work on genetics with Edgar as well as on his lectures at Hughes Company about biology, organic chemistry and microbiology. This will be addressed in Section III, after a section devoted to a quick historical reconstruction about the birth of molecular biology, which was properly the field of interest of Feynman. Finally, in Section IV, conclusions and outlook will be presented.

## 2. THE "PHAGE GROUP" AND THE BIRTH OF MOLECULAR BIOLOGY

Molecular biology came into play as new research paradigm during the three decades ranging from 1930 to the late 1950s, characterized by a huge effort to understand the secret of life, whose main result was the discovery of the self-replicating mechanisms of DNA and the explanation of its working principle: the information coding. That triggered further developments which contributed to the development of genetic engineering. United States – and in particular two institutions – played a dominant role in this respect: the Rockefeller Foundation, who launched and supported an intensive biology program, and Caltech, who carried out Rockefeller's project and became the top international research and training center in molecular biology (see for instance (Kay 1993; Keller 2000; Poon 2001; Keller 2002; de Chadarevian 2002; Joaquim, Freire and El-Hani 2015) and references therein).

### 2.1 Building up a new science

The term "molecular biology" was coined in 1938 by Warren Weaver, the director of the Rockefeller Foundation's natural science division. It well captured in its meaning the content of the Foundation's program and, in general, the main features of the new science: 1) the focus on unifying life phenomena common to all living organisms; 2) the use of simple biological systems – such as bacteria and viruses – as phenomenological probes or conceptual models; 3) the search for ultimate physicochemical laws governing all living phenomena; 4) its interdisciplinary nature in borrowing concepts and methods from different fields such as physics, mathematics, chemistry, genetics, microbiology, immunology and physiology; 5) a domain of investigation ranging from $10^6$ to $10^7$ cm; and 6) the use of new and more sophisticated instrumentation and techniques.

In the years from 1930 to the late 1950s, the significant role of Rockefeller Foundation in shaping life science and, in particular, molecular biology was the result of a variety of different factors, which run from huge investments to a smart scientific policy consisting in creating and supporting mechanisms of interdisciplinary cooperation through networks of grants and fellowships as well as promoting a strongly project-oriented research. Rockefeller projects and university research programs became soon deeply interconnected, and the research in molecular biology grew up as a result of an overall strategy, based on an interdisciplinary cooperation and the so-called group projects. At the same time, the quest for new and sophisticated experimental equipment (and, as a consequence, for larger laboratories to house such equipment) triggered the development of new technologies, also demanding a close interplay between biology, physics and engineering.

Among the institutions which received a large amount of grants for carrying out projects in molecular biology we find the University of Chicago and Caltech, considered by the Foundation as the most promising centers for developing the new cutting-edge research programs. But, at variance with Chicago, Caltech's biology program had, as a unique feature, a sharp departure from the traditional point of view in biology: the aim was, indeed, to build up a new science, mainly based on a fruitful interplay with engineering and physics. This is testified by the definition of new curricula for undergraduate and graduate studies in biology, which featured a strong training in physical sciences. As a consequence, Caltech soon became a primary research center in molecular biology.

## 2.2 Delbruck and the phage group

The physicist Max Delbruck was one of the founding fathers of molecular biology, who worked at Caltech and built up influential research groups. He was the first to establish successful links between physics, genetics and mathematics by creating the "phage group" in the late 1930s. The leitmotif of his research program was an emphasis on bacterial viruses (or bacteriophages), taken as model system for gene action, in this way introducing a new working approach in molecular biology. However, a key activity of the group, which enabled Delbruck to keep close contact with the scientific community on genetics – and boosted his career – was the organization of summer symposia taking place every year in Cold Spring Harbor. It was just in the 1941 Symposium that Delbruck, for the first time, presented a paper on protein chemistry: it focused on a possible analogy between self-replication and enzymatic autocatalytic reaction, and recognized an enzyme-like protein as the active hereditary component of chromosomes. At the same time, he started his life-long collaboration with Salvador Luria.

An electron microscope enabled Delbruck in 1943 to observe a bacteriophage and to elucidate its structure: a tadpole-shaped or sperm-like organism with distinct head and tails. That suggested him the close analogy between phage penetration of bacteria and the interaction of sperm with the egg, which added new evidence to the possible relation between the specificity of proteins involved in genetic replication and the specificity involved in the formation of antibodies.

In 1945 an annual phage course was organized in Cold Spring Harbor, which would be held till late 1960s, while the number of researchers on such topic quickly grew up. The course primarily dealt with borderline problems in biology, chemistry and physics, and Delbruck required a strong mathematical background to the participating students, along with a knowledge of basic laboratory techniques. The close contact with Niels Bohr, as well as the inspiration from the book *What is life?* by Erwin Schrödinger (1944), contributed a lot to Delbruck successful career and promoted molecular biology as a line of research for physicists.

A further byproduct was the broadening of the scientific interests of the phage school, as testified by Delbruck and Bailey (1946) and Alfred D. Hershey (1946) papers presented at 1946 Cold Spring Harbor Symposium.

Here a clear evidence for a complicated genetic behavior by bacterial viruses was provided for the first time, showing that they undergo mutations mainly during their intracellular existence. In other words, a microorganism was a complex reproductive system able to transmit specific genetic factors, which could be identified with nucleic acids. Further work by Hershey and Delbruck showed that a genetic map of a phage could be constructed, in whole analogy with Drosophila maps, paving the way for later reconstruction of a fine-structure map of the phage genome. As a result, the classical concept of gene changed because of the separation between units of recombination, mutation and function, while new experiments were designed in order to assess the primacy of DNA during replication and mutation in phage.

Delbruck's research carried out between 1940 and 1946 led him to play a primary role in the fields of genetics and microbiology, and raised the interest of many research institutions. He made his choice in January 1947, joining Caltech as a full professor, and there set up a permanent research and training center on phage, with a laboratory endowed with novel and advanced experimental technologies. In this respect, new tools such as the radioisotope tracer began to come into play in molecular biology and revealed themselves very powerful. Indeed, starting in 1947, radioisotopes were used in phage studies, as reported in the contributions presented at the yearly Cold Spring Harbor Symposium.

**2.3. Finding the replication mechanism**

Caltech group attracted a lot of scientists in those years, contributing in this way to set up the basic pillars in molecular biology, which culminated with the discovery of DNA double helix structure by James D. Watson and Francis H.C. Crick in 1953.

Among the main contributions to the development of the new science, which took place since 1947 under the influence of Delbruck's group at Caltech, we have to quote the multiplicity reactivation phenomenon by Luria, which deals with the genetic exchange of undamaged parts between ultraviolet irradiated phage particles during the process of absorption to the same host bacterium. Further investigations on x-rays damage on phages and their patterns of recombination were the subject of Ph.D thesis by Watson under the supervision of Luria and Delbruck, the perspective being the search for the general relationships between structure and function in viruses.

Subsequent experimental findings by Doermann (1948) about the phage life cycle triggered further investigations by Delbruck, Hershey and Luria about segregation and recombination of viral genetic material during the vegetative phase, while in 1950 Lwoff added new pieces of information about the mechanisms of replication and mutation in bacteria (Lwoff 1966). The acquired knowledge of the phage life cycle in the bacterial cell was soon extended to different animal and human viruses by Renato Dulbecco, who joined Delbruck's laboratory in 1950. Following a suggestion by Delbruck, he succeeded in developing a method for the growth of animal cells able to produce viruses in culture dishes. As a result, a novel and reliable plaque assay for viruses was established in whole analogy with the phage case (Dulbecco 1966), paving the way to the development of molecular virology.

All these findings pointed clearly toward the key role played by nucleic acids in the replication and mutation in phage, but this idea remained unexplored till 1953, while protein research being fully pursued by George Beadle and Linus Pauling. In this context, the success of Pauling's project on sickle cell anemia (Pauling *et al.* 1949) confirmed the role of giant protein molecules in all physiological functions and pointed out how the etiology of disease could be found at the molecular level. In other words, Pauling's study on sickle cell anemia was a first example of a molecular disease, the manufacture of abnormal sickle cell hemoglobin being controlled by a particular gene, in this way validating the molecular vision of life.

On the basis of these results, in January 1950 Beadle and Pauling requested funding from the Rockefeller Foundation to build up a laboratory of medical chemistry at Caltech, whose mission would be to bring together in a stimulating scientific environment biologists, chemists, physicists and experts in medicine in order to understand the chemical processes underlying biological systems. But the skepticism of the Foundation's officers about their proposal led them to give up soon and redraw their attention and energies to protein structure.

Indeed, a relevant scientific achievement by Pauling in those years was the construction of the physical model of alpha-keratin, a task pursued relying strongly on the building of molecular models, which was the hallmark of Pauling's research activity, well known as molecular architecture (Bernal 1968; Corey and Pauling 1953). The alpha helix, with the pitch of the turn occurring every 3:7 amino acid residues, revealed a strong departure from known protein structures, being a helix of peptides with an irrational, aperiodic structure. But the issue of finding the auto-replication mechanism of such a structure remained still unknown. Proteins didn't solve the main problems in biology, a possible answer having to be found in nucleic acids.

A scientific revolution was about to happen, in which life appeared to be ruled by a new giant molecule, the self-replicating DNA spiral. Its double-helix structure sustained by a complementary pairing of purines to pyrimidines, elucidated by Watson and Crick (1953), suggested a possible copying mechanism of the genetic material. Delbruck's reactions were enthusiastic, and Watson was invited to give a talk at the 1953 Cold Spring Harbor Symposium on viruses. It is clear that the shift of the molecular vision of life, from the protein paradigm to the new DNA based one, determined the beginning of a new era in molecular biology, centered essentially on genetics and cytology, whose main achievements were the discovery of DNA replication mechanism together with the role of DNA polymerase as a catalyst, and the development of the idea of a genetic code as a solution to the problem of heterocatalysis.

Delbruck's phage group and the funding policy carried out by Rockefeller Foundation revealed to be a fundamental contribution to the birth of molecular biology, and Caltech began a primary research center in this respect till the end of 1960s.

## 3. FEYNMAN AT WORK IN BIOLOGY

Being the world center of molecular biology research in 1950s-1960s, all the leaders in the field sooner or later would visit the biology department at Caltech. Feynman too, who often visited Delbruck, often attended seminars given by these visitors (Mehra 1994), and being a frequent visitor of the biology department, he was able to meet for example Dulbecco and Seymour Benzer, who later would give colorful accounts and anecdotes about their interaction with him (Dulbecco 2010; Weiner 1999).

### 3.1 Plus and minus classes: Feynman at Caltech

At some time, Feynman realized that he might like to do some work in biology, and then Delbruck sent him Robert S. Edgar – Delbruck's postdoc at the time – who was carrying on bacteriophage research, which Delbruck was losing interest into. As a task, he was given to work on back-mutations, i.e. mutations appearing to restore a mutant gene to its normal state. It is important to notice that back-mutations do not always bring back to exactly the starting point. His work follows previous studies by Benzer, who first recognized the uniqueness of *rII* mutants, namely, their inability to form plaques on Escherichia Coli *K12*. According to him, this property could be useful to analyze the nature of genes, because it allows a small fraction of wild-type recombinants from crosses to be easily enumerated. In this way, it is possible to study the detailed genetic fine structure of the *rII* region (Benzer 1955; 1961). Benzer was able to genetically map a huge number of mutations in the *rII* gene, and that allowed him to understand two main features about genes: the sequence of a gene is linear and the smallest units of recombination is between two adjacent DNA base pairs.

Feynman's work consisted in mapping a reasonably large number of *rII* markers in a second phage strain, the T4D one (Edgar *et al.* 1962). By analyzing back-mutants that were evidently not completely normal, he realized that such back-mutants had both the $r_{43}$ mutation and a second mutation that somewhat enhanced its effects. Such mutations – which we may call "suppressors" – had by themselves quite a strong effect, similar to that of $r_{43}$. However, when combined with $r_{43}$, they brought back the phage close to the starting, normal state. Feynman also showed that different suppressors, when combined between them, do not produce mutual suppression, but rather they appear to suppress only the $r_{43}$ mutation: the former were shown to be located near the latter.

By studying back-mutations of suppressors, Feynman found that they were due to new suppressors similar to the $r_{43}$ mutation, which were referred to as plus and minus mutations. Combination of a plus and a minus mutation brings the phage almost back to its normal state. Such a picture was confirmed by Crick *et al.* (1961) in the famous paper where the genetic code was unveiled, showing that each amino acid in the protein synthesis corresponds to three nucleotides. Feynman went close to such a finding, but did not realize the importance of what he had uncovered. In Benzer's words:

> "He had discovered something without realizing it. [...] It was related to the later discovery by Crick and Brenner, using the *rII* mutants. This had to do with the nature of the genetic code. [...] It was something under his nose, and its significance was just not apparent at that time" (Benzer 2002).

What Feynman was missing – while known to Crick et al. – was that the plus and minus mutations corresponded to additions and deletions of nucleotides, respectively. Also, he did not understand that the number three was peculiar, and to be identified with the coding ratio; this was famously discovered by Crick and coworkers in the mentioned paper (Crick *et al.* 1961).

### 3.2 A course on biology, organic chemistry and microbiology: Feynman at Hughes

In the fall of 1955, Robert Hellwarth, who joined Caltech Physics Department as a research fellow, together with Frank Vernon, an engineering research student working at Aerospace Corporation, drew Feynman's interests on more applied research topics. In 1956 Hellwarth moved to Hughes Aircraft Company and arranged for Feynman to give there lectures for scientists, engineers and technicians on subjects of mutual interest. Feynman continued lecturing regularly at Hughes for many years on a variety of topics, including in particular molecular biology.

The lectures went on regularly until the end of the 1970s, reserved to the employees of the Company, but unfortunately there was no audio or video recording systems, so that we can rely only on notes taken by the attendees. In particular, notes for the Statistical Mechanics lectures of 1961 were taken by R. Kikuchi and H.A. Feiveson; these notes were later published in the now famous book Statistical Mechanics: a set of lectures (Feynman 1972). Other sets of notes were taken by J.T. Neer, who later made them freely available on the web (Feynman 1970). The other lectures apparently went unrecorded. The notes taken by Neer include lectures given by Feynman from October 1966 to June 1971 about the following topics:

1. October 1966 - June 1967:
   Astronomy, Astrophysics, Cosmology;
2. October 1967 - June 1968:
   Electrostatics, Electrodynamics, Matter-Waves Interacting, Relativity;
3. July 1968 - June 1969:
   Matter-Wave Interacting Continued, Introduction to Quantum Mechanics, Scattering Theory, Perturbation Theory, Methods & Problems in QED;
4. October1969 - May 1970:
   Biology, Organic Chemistry and Microbiology;
5. October 1970 - June 1971:
   Mathematical Methods in Engineering & Physics.

These sets of notes were only slightly edited, therefore are a good example of Feynman in action. This is especially intriguing for the first and the fourth sets, which illustrate Feynman dealing with fields outside his main research, using lectures as a mean to enter a subject he was interested in. Now we will focus on the fourth set of lectures (Feynman 1970), i.e. those lectures concerning molecular biology. As discussed above, Feynman was not new to biology in 1969, having worked previously in a biology lab for one year, but, being an outsider, he found the material challenging and time consuming. As a result, this set of lectures is considerably shorter than the other sets and, moreover, the lectures ended earlier than expected, Feynman being more and more involved in that period with the development of his parton theory (Feynman 1969).

The lectures highlight quite a standard course on organic chemistry, biomolecules, genetics, and microbiology; the topics covered are reported in Table I. However, some considerations are present here and there that betray his being a physicist.

In the introduction, Feynman noted that, unlike physics and chemistry, biology lacks a basic foundation of fundamental laws, developed by theory and proven by experiments. Lacking such a guiding principle, he organized the material according to scale, ranging from the molecular level to more and more complex systems, up to ecology, i.e. the study of many complex biological system interacting in a closed environment. Feynman was thus naturally led to the molecular biology approach, according to which “the chemical constituents react according to known chemical and physical laws in a manner which can account for life" (Feynman 1970). He was as well convinced that “he could derive all of the properties of living things from the quantum mechanics of the carbon atom" (Bridges 2004).

After the introduction, Feynman began a brief survey of the essentials of organic chemistry: hydrocarbons, functional groups, alcohols, carbonyl compounds, esters, chiral molecules. After that he switched to biochemistry, i.e. biomolecules and metabolic pathways, then he discussed sugars and cellular energy production (photosynthesis, Krebs cycle). After that he continued with other biomolecules, i.e. fats, amino acids and proteins, discussing in detail the structure of the latter, going from the alpha-helix to globular proteins, highlighting the role of hemoglobin and myoglobin.

The course then turned to molecular biology, namely the structure of nucleotides and of the nucleic acids, discussing DNA reproduction, the genetic code, protein synthesis and mutations. In the last part of the course the focus shifted, as announced, to more complex systems such as the retina, antibodies, cell differentiations, nerve cell growth and social amoebas. As said above, however, the course was interrupted by Feynman earlier than expected, so that no further discussion on microbiology is present, nor on the planned ecology section.

## 4. CONCLUDING REMARKS

In the late 1950s, Feynman was deeply involved with a number of physics researches, where he actually gave important contributions. In addition to studies on quantum gravity (Feynman 1963) and, especially, to his well-known results about the V-A (vector-axial) character of weak interactions, the two-component spinor formulation of the Dirac equation (Gell-Mann and Feynman 1958) and the density matrix approach to polaron theory in solid state physics (Feynman *et al.* 1962) (just to mention some examples), his own character led him to devote himself also to calculations of the tracking of artificial satellite Explorer II at the Jet Propulsion Laboratory (Mehra 1994) or even to pedagogical work (as his most famous Lectures on Physics (Feynamn, Leighton and Sands 2005), for example) and popularization (about nanotechnology (Feynman 1960), just to mention one) issues. In any of these topics Feynman excelled but, in our opinion, rather than being the manifestation of a genius at work, this is more appropriately the epiphany of his peculiar curiosity, which brought him to be interested also in possible algebraic manipulations performed by computers or other similar, apparently strange things for a well pictured theoretical physicist.

However, it is probably Feynman's unexpected involvement in biological issues that better highlights what truly lies behind his curiosity. Indeed, it is somewhat apparent from what was discussed above in the present paper that it was not properly the satisfaction for testing one's own abilities in getting some important result in even different fields of research (even for social utility and not for egoistic purposes) that drove Feynman's curiosity, but rather what we may call the challenge to understand Nature in all its different facets.

This line of reasoning and doing was already well apparent to be in action in the framework of physics, the original field of Feynman, but the deep roots of its foundations are much more appreciated in the framework of biology, a novel field of Feynman's interests. Future studies in this direction will probably reveal other intriguing features of one of the most interesting minds of our times.

---

| Feynman Hughes Lectures on:<br>**Biology, Organic Chemistry and Microbiology** |
|---|
| Notes taken by: **John T. Neer**<br>Date: **October 1969 - May 1970** |
| - Introduction to the Course on Biology |
| - Organic Chemistry<br>- Introduction<br>- Alkenes<br>- Alcohols<br>- Carbonyl compounds<br>- Asymmetric carbon compounds<br>- Sugars |
| - Biochemistry<br>- Introduction<br>- Carbohydrate metabolism<br>- Photosynthesis<br>Fixation of carbon<br>- Substances of Life<br>Fats<br>Proteins<br>- Protein structure<br>The pleated sheets and alpha-helix<br>The alpha-helix<br>alpha-keratin<br>Collagen<br>Globular proteins<br>Cytochrome C<br>- The structure of nucleic acid and polymers. DNA and RNA<br>Introduction<br>DNA<br>DNA reproduction<br>Genetic code<br>Protein<br>synthesis<br>Mutation |
| - Genetics<br>- Meiosis<br>- Sex determination<br>- Control<br>- Allostery<br>- Production of various amino acids<br>- mRNA<br>- Control of recrealing of DNA<br>- How did all begin<br>- Antibody reaction<br>- Fertilization. Cell division<br>- Cell differentiation<br>- Animal metamorphism<br>- Social amoeba |

TABLE I: Topics of Feynman's course on Biology, Organic Chemistry and Microbiology, as deduced by Neer notes of the corresponding Hughes Lectures.